\documentclass[a4,journal]{IEEEtran} 
\usepackage{amsmath,amsfonts,amssymb}
\usepackage{cite}
\usepackage[ruled,linesnumbered]{algorithm2e}
\usepackage{array}
\usepackage[caption=false,font=normalsize,labelfont=sf,textfont=sf]{subfig}
\usepackage{textcomp}
\usepackage{stfloats}
\usepackage{url}
\usepackage{verbatim}
\usepackage{graphicx}
\usepackage{cite}
\usepackage[export]{adjustbox}
\usepackage{cancel,xcolor}

\usepackage{glossaries}

\newacronym{TkBD}{TkBD}{track-before-detect}

\newacronym{DBTk}{DBTk}{detect-before-track}

\newacronym{VAR}{VAR}{vector auto-regressive}

\newacronym{SNR}{SNR}{signal-to-noise ratio}

\newacronym{BTR}{BTR}{bearing time record}

\newacronym{CFAR}{CFAR}{constant false alarm rate}

\newacronym{PMBM}{PMBM}{Poisson multi-Bernoulli mixture}

\newacronym{PDF}{PDF}{probability density function}

\newacronym{DOA}{DOA}{direction of arrival}

\newacronym{OSPA}{OSPA}{optimal subpattern assignment}

\newacronym{LOFAR}{LOFAR}{low frequency analysis and recording}

\newacronym{BRFS}{BRFS}{Bernoulli random finite set}

\usepackage[utf8]{inputenc}
\usepackage{pgfplots}
\usepgfplotslibrary{groupplots,dateplot}
\usetikzlibrary{patterns,shapes.arrows}
\pgfplotsset{compat=newest}

\usepackage{comment}

\newcommand{\norm}[1]{\left\lVert#1\right\rVert}
\renewcommand{\bf}[1]{\mathbf{#1}}

\newcommand{\dBB}{{(\mathrm{dB})}}

\newcommand{\likelihoodratio}{L}
\newcommand{\likelihood}{\ell}
\DeclareMathOperator*{\argmin}{arg\,min}

\newcommand*{\expect}{\mathsf{E}}
\usepackage{booktabs}
\usepackage{siunitx}

\DeclareMathAlphabet\mathbfcal{OMS}{cmsy}{b}{n}

\usepackage{bm}
\usepackage{bbm}
\renewcommand{\bm}[1]{\vec{#1}}

\usepackage{makecell}
\usepackage{inconsolata}

\begin{document}
\title{Broadband Passive Sonar Track-Before-Detect Using Raw Acoustic Data}

\author{
	Daniel Bossér (Student Member IEEE), Magnus Lundberg Nordenvaad, Gustaf Hendeby (Senior Member IEEE), Isaac Skog (Senior Member IEEE)
\thanks{This work has been founded by the research organization Zenith (project id. 20.03) and SecurityLink.}
\thanks{Daniel Bossér and Gustaf Hendeby are with Dept. of Electrical Engineering,
Linköping University, (e-mail: daniel.bosser@liu.se; gustaf.hendeby@liu.se).}%
\thanks{
Magnus Lundberg Nordenvaad is with the Div. of Underwater Technology, Swedish Defence
	Research Agency (FOI), Kista, Sweden (e-mail: magnus.lundberg@foi.se)
}
\thanks{Isaac Skog is with Dept. of Electrical Engineering and Computer Science, KTH Royal Institute of Technology,
	and the Div. of Underwater Technology, Swedish Defence
Research Agency (FOI), Kista, Sweden (e-mail: skog@kth.se)}}

\markboth{}%
{Shell \MakeLowercase{\textit{et al.}}: A Sample Article Using IEEEtran.cls for IEEE Journals}


\maketitle

\begin{abstract}
This article concerns the challenge of reliable broadband passive sonar target detection and tracking in complex acoustic environments. Addressing this challenge is becoming increasingly crucial for safeguarding underwater infrastructure, monitoring marine life, and providing defense during seabed warfare. To that end, a solution is proposed based on a vector-autoregressive model for the ambient noise and a heavy-tailed statistical model for the distribution of the raw hydrophone data. These models are integrated into a Bernoulli track-before-detect (TkBD) filter that estimates the probability of target existence, target bearing, and signal-to-noise ratio (SNR). The proposed solution is evaluated on both simulated and real-world data, demonstrating the effectiveness of the proposed ambient noise modeling and the statistical model for the raw hydrophone data samples to obtain early target detection and robust target tracking. The simulations show that the SNR at which the target can be detected is reduced by \qty{4}{\dB} compared to when using the standard constant false alarm rate detector-based tracker. Further, the test with real-world data shows that the proposed solution increases the target detection distance from \qty{250}{\meter} to  \qty{390}{\meter}. The presented results illustrate that the TkBD technology, in combination with data-driven ambient noise modeling and heavy-tailed statistical signal models, can enable reliable broadband passive sonar target detection and tracking in complex acoustic environments and lower the SNR required to detect and track targets.		
\end{abstract}

\begin{IEEEkeywords}
	Underwater passive survelliance, target tracking, array signal processing
\end{IEEEkeywords}

\section{Introduction }


A significant portion of today’s critical infrastructure is located underwater.
This includes gas pipelines, power transmission lines, and communication
cables, which are essential in modern society. Due to their remote
locations and strategic importance, these assets are vulnerable to damage and
sabotage. Monitoring these infrastructures is crucial, especially during times
of conflict, when targeted attacks on them could have severe consequences
\cite{Soldi2023}. This has led to the emergence of a new domain of conflict
on the world’s seafloors, known as seabed warfare, which necessitates
the development of advanced countermeasures \cite{Bashfield2024,
alleslev2019nato}.

Passive sonar surveillance is essential for discreetly monitoring underwater
infrastructure. This technique enables the detection of submarines and other
underwater vehicles without disclosing the location of the sonar system.
Moreover, passive sonar does not introduce noise pollution in the ocean, benefiting marine
life \cite{Dolman2011, Parsons2017}, while allowing for continuous, unobtrusive
monitoring of the underwater environment.
However, compared to active sonar, passive sonar systems typically operate at a
lower \gls{SNR} and are more susceptible to ambient noise. This requires
them to have higher sensitivity and use more complex noise models.


Passive surveillance has historically relied on a combination of signal
processing techniques such as \gls{LOFAR}, beamforming, and \gls{BTR}
analysis. These methods are often manually operated by human sonar operators,
who may also listen to the sounds. However, relying on human expertise
is costly and resource-limited, making large-scale monitoring challenging. Energy-based detectors, such as the
\gls{CFAR} detector, have been employed to automate the surveillance process. These detectors
output a set of bearings at each time step, corresponding to potential
target detections. When a target is present, some detections may over time form a track, which can be identified using target tracking methods such as
multiple hypothesis tracking \cite{Blackman2004} or more recent
approaches such as the \gls{PMBM} filter \cite{GarciaFernandez2018}.
Applications of these methods to underwater surveillance have been explored in
\cite{Fortmann1983, Li2021}. However, a prerequisite for successful detection
is that the \gls{SNR} is sufficient to exceed the detection threshold.

\begin{figure}
	\centering
	\subfloat[No target]{\includegraphics[trim = {0.5cm 0.55cm 0.5cm 0.1cm}, clip, width = 0.98\linewidth]{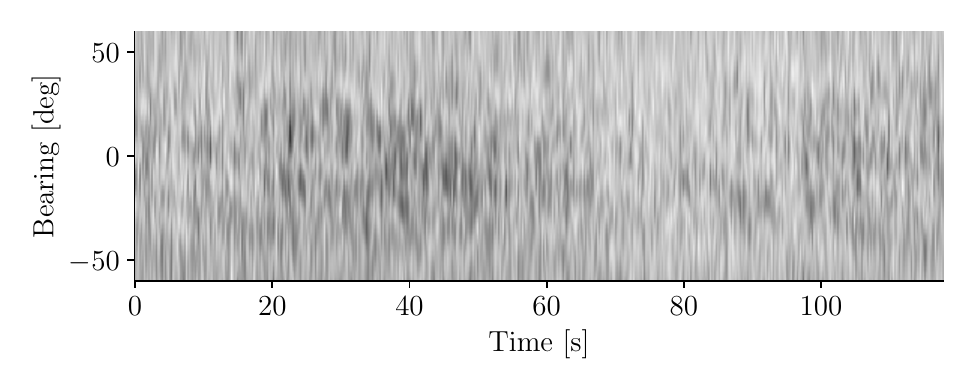}}\\
	\subfloat[With target]{\includegraphics[trim = {0.5cm 0.55cm 0.5cm 0.5cm}, clip, width = \linewidth]{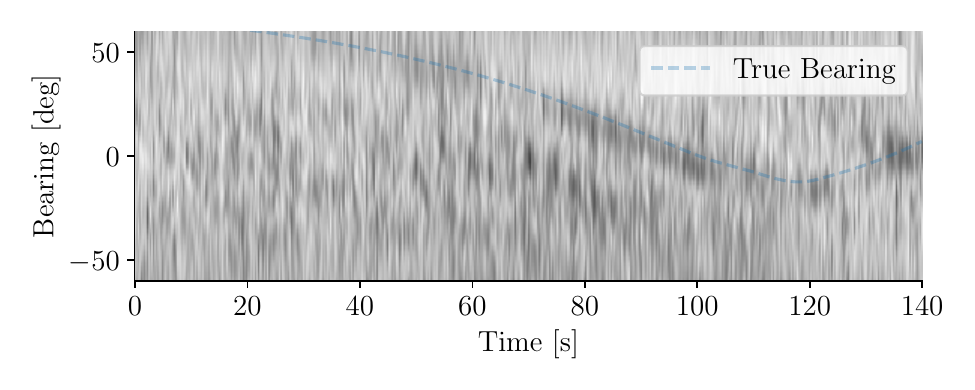}}
	\caption{
		\label{fig:btr-early}
		 BTRs from data collected during a sea trial. The BTR values were calculated using a conventional delay-and-sum beamformer. The upper and lower plots show the BTR without and with a target present. From the plots, it is clear that the temporal and spatially correlated ambient noise causes disturbances that are hard to distinguish from the actual target.}
\end{figure}

One way to increase the performance in poor \gls{SNR} is by using the \gls{TkBD}
tracking strategy \cite[p. 239]{Ristic2003Beyond}. In this approach, the target
detection occurs much later in the signal processing chain, after constructing the potential
track, hence the name. Thus, target detection and tracking are done jointly. The
major benefit of this method is that no information is discarded in the
detection process, allowing for a longer integration time of the raw data
before the decision is made, consequently lowering the \gls{SNR} requirement.
Theoretically, a performance gain of approximately \qty{6}{dB} is possible
\cite[p. 318]{Mallick2012Tracking}. However, \gls{TkBD} makes the
tracker more sensitive to signal and noise modeling errors, which is why the application of \gls{TkBD} to the
underwater passive sonar problem has been so challenging. Consider the \gls{BTR}
in Fig.~\ref{fig:btr-early}a,  which shows the received signal energy in
different bearings over time. From
Fig.~\ref{fig:btr-early}a, it is evident that the signal
energy varies greatly over time and bearing, implying that the received signal
is spatiotemporally correlated. These correlations are difficult to
discern from the signal components produced by a target, as seen in
Fig.~\ref{fig:btr-early}b. Another challenge is that ambient noise is known
to exhibit a heavy-tailed distribution \cite[p.~403]{Abraham2019}. Previous work
has attempted to model the acoustic samples using alpha-stable distributions
\cite{Song2019} or Gaussian mixture models \cite{Zhang2020a} to address this
issue, but this has not been done in the context of target tracking.

Different approaches utilizing \gls{TkBD} for passive sonar are explored in \cite{Xu2017, Peng2019, Northardt2019, Yi2019}. Many of these methods rely on modeling signal energy post-beamforming as their measurements and only consider narrowband signals. Some approaches have attempted to address the challenges posed by spatiotemporal correlations. For instance, the work in \cite{Northardt2019} averages the signal energy post-beamforming in different bearing bins, thereby reducing the impact of the temporal energy variations. Similarly, the post-beamforming energy is also used in \cite{Yi2019}. By comparing the beamforming energy in the presence and absence of a target, the authors of \cite{Yi2019} fit a probability distribution to each bearing bin. Although this approach adapts to spatial variations in ambient noise, the fitted distributions are time-invariant, meaning that temporal energy variations are not accounted for. Likewise, the work in \cite{Yocom2011} and \cite{Bosser2023} applies models that assume noise interferers at specific bearings that emit independent, constant power signals. None of these models handle the temporal variation in the ambient noise.

\subsection{Contributions}

This article builds upon and extends the work presented in~\cite{Bosser2022, Bosser2023}, which explored the
possibility of sample-level source and ambient sound modeling in a passive sonar \gls{TkBD} application, circumventing the challenges of developing accurate statistical models for the signal after beamforming. In~\cite{Bosser2023}, it was observed that spatiotemporal correlations and other modeling errors make the tracker prone to initiating false tracks. This article proposes a solution to this challenge by adding a new data preprocessing step and a new measurement model for the data samples. In summary, the contributions of this work are:
\begin{enumerate}
	\item a \gls{VAR} model for modeling of the spatially and  temporally correlated ambient noise; 	
	\item a measurement model for wideband signals with a heavy-tailed distribution; and
	\item an experimental evaluation of the proposed models within a bearing-only \gls{TkBD} framework.
\end{enumerate}
\textbf{Reproducible research:} The code and data used to reproduce the presented results can be downloaded at \texttt{https://gitlab.liu.se/coast/tkbd\_using\_raw\_data}.

\section{Target Tracking and Bayesian Estimation}
\label{sec:target_tracking}

\newcommand{\ti}{k}

The objective of target detection and tracking is to, given a set of measurements
$z_{1:\ti} = \{z_1, z_{2}, \dots, z_\ti\}$
collected at discrete time instances $1, 2, \dots, \ti$, determine if a target is present
and, if so, estimate its current state.
Commonly, the measurements $z_{1:\ti}$ are a set of detections, but in
\gls{TkBD} applications, they may instead be intensity measurements, images,
beamforming outputs, or, as this article proposes, a set of hydrophone samples.

To simultaneously describe the target state and the probability of its existence, the
target is modeled using a \gls{BRFS}. To that end, let
$q_\ti$ denote the probability of existence, and $x_\ti$ denote the state of the target, given that the target exists.
The \gls{BRFS} $X_\ti$ jointly describes these components with the finite set statistics (FISST) \gls{PDF} \cite{Mahler2007}
\begin{align}
	\label{eq:berrfs_pdf}
	f_\ti(X) = \begin{cases}
		1 - q_\ti \quad & \mathrm{if} \, X = \emptyset, \\
	q_\ti s_\ti(x) \quad & \mathrm{if} \, X = \{x\},
	\end{cases}
\end{align}
where $s_\ti(x)$ is the \gls{PDF} of state $x_\ti$.
The posterior \gls{PDF}
\begin{equation}
	\label{eq:berrfs_posterior}
	f_{k|k'}(X) \equiv f_k(X|z_{1:k'}) = \begin{cases}
		1 - q_{k|k'} \, & \mathrm{if} \, X = \emptyset, \\
		q_{k|k'} s_{k|k'}(x) \, & \mathrm{if} \, X = \{x\},
	\end{cases}
\end{equation}
can be calculated using the Bernoulli filter \cite{Ristic2013}. Here,
$q_{k|k'}$ and $s_{k|k'}(x)$ denote the posterior
probability of existence and posterior state distribution at time instant $k$ given measurements up to time instant $k'$.
Given the posterior \gls{PDF} $f_{\ti - 1|\ti-1}(X)$,
the Bernoulli filter recursions for calculating the posterior distribution $f_{\ti|\ti}(X)$
are given by the time and measurement updates \cite{Ristic2013}
\newcommand{\pb}{p_\mathrm{b}}
\newcommand{\pss}{p_\mathrm{s}}
\begin{subequations}
\begin{align}
	q_{\ti|\ti - 1}   & =  p_\mathrm{b} (1 - q_{\ti - 1|\ti - 1}) + p_\mathrm{s} q_{\ti - 1| \ti - 1}, \\
\begin{split}
	s_{\ti|\ti - 1}(x)  & = \frac{\pb (1 - q_{\ti - 1| \ti - 1} b_{\ti | \ti -  1}(x))}{q_{\ti | \ti - 1}} \\
			    & + \frac{\pss q_{\ti - 1 | \ti - 1} \int \pi_{\ti|\ti - 1}(x|x') s_{\ti - 1| \ti - 1}(x') \, dx' }{q_{\ti | \ti - 1}},
\end{split}
\end{align}
\end{subequations}
and
\begin{subequations}
	\begin{align}
		q_{\ti|\ti} = & \frac{q_{\ti|\ti - 1} \int \likelihoodratio(z_\ti|x) s_{\ti | \ti - 1}(x) \, dx }{1 - q_{\ti|\ti - 1} + q_{\ti|\ti - 1} \int \likelihoodratio (z_\ti|x) s_{\ti | \ti - 1}(x) \, dx }, \\
		\label{eq:ordinary_bayesian_filter_update}
		s_{\ti|\ti}(x) = & \frac{\likelihoodratio(z_\ti|x) s_{\ti|\ti - 1}(x)}
		{\int \likelihoodratio(z_\ti|x) s_{\ti|\ti - 1}(x) \, dx},
	\end{align}
\end{subequations}
respectively. Here, $\pb$ and $\pss$ are the probability of target birth
and survival between time steps, respectively. Further, $b_{\ti|\ti-1}(x)$ denotes the \gls{PDF} for states of newly born targets. Moreover, $\pi_{\ti|\ti - 1}(x|x')$ is a
\gls{PDF} of the state $x$ conditioned on the state $x'$, describing the target motion.
Finally, $\likelihoodratio(z_\ti|x)$ is the conditional likelihood ratio of given the target state
$x$. This article uses the particle filter in \cite{Ristic2013} to implement and execute these filter recursions.

To execute the filter recursion, one must specify models for the target dynamics, target birth
and survival probability, and measurement likelihood. This article focuses on how the measurement likelihood should be modeled in
a broadband passive sonar system. Next, the pros and cons of the commonly used CFAR
likelihood model will be reviewed, and a new likelihood model that addresses some
of the cons of existing measurement models will be proposed.

\section{Measurement Models}
\label{sec:measurement_model_new}

The section describes how the relationship between the measurements $z_\ti$ and the target state $x_\ti$, and the associated
likelihood ratio function $\likelihoodratio(z_\ti|x_\ti)$, is modeled. First, the traditional case when $z_\ti$ consists of
detections is described. Second, a new measurement model for when $z_\ti$ consists of raw hydrophone samples is presented.

\subsection{Spatial Signal Processing and Beamforming}
\newcommand{\ssi}{n}
Consider a hydrophone array consisting of $M$ hydrophones. Let $y^{(m)}_\ssi$ denote sample $\ssi$ from hydrophone $m$ in the array.
Define \begin{equation}
\bm{y}_\ssi = \begin{bmatrix}
	y_\ssi^{(0)} & \dots & y_\ssi^{(M - 1)}
\end{bmatrix}^\top \in \mathbb{R}^{M},
\end{equation}
as the collection of samples from all the $M$ hydrophones. Further, let
\begin{equation}
	\mathrm{y}_\ti = \begin{bmatrix} \bm{y}_{(\ti-1)N+1}^\top & \dots & \bm{y}_{\ti N}^\top \end{bmatrix}^\top \in \mathbb{R}^{N M},
\end{equation}
be a batch of $N$ hydrophone samples.

If the target is in the far field and emits a broadband signal $\mathrm{s}_\ti \in \mathbb{R}^{N}$, the batch of hydrophone samples $\mathrm{y}_\ti$ can be modeled as
\begin{equation}
	\mathrm{y}_\ti = \mathrm{H}(\psi) \mathrm{s}_\ti + \mathrm{e}_\ti,
\end{equation}
where $\mathrm{e}_\ti$ is given by
\begin{subequations}
\begin{align}
	\mathrm{e}_\ti & = \begin{bmatrix} \bm{e}_{(\ti-1)N+1}^\top & \dots & \bm{e}_{\ti N}^\top\end{bmatrix}^\top \in \mathbb{R}^{NM}, \label{eq:e_batch} \\
	\bm{e}_\ssi & = \begin{bmatrix}
		e_\ssi^{(0)} & \dots & e_\ssi^{(M - 1)}
	\end{bmatrix}^\top\in \mathbb{R}^{M},
\end{align}
\end{subequations}
and $e_\ssi^{(m)}$ is the measurement noise of hydrophone $m$ at time instant $n$. Furthermore, $\mathrm{H}(\psi)$ is a fractional
delay filter matrix, given by \cite{Pei2012}
\begin{align}
	\mathrm{H}(\psi) & = \begin{bmatrix} H_0^\top(\psi) & \dots & H_{M - 1}^\top(\psi) \end{bmatrix}^\top \in \mathbb{R}^{NM \times N},
\end{align}
where
\begin{subequations}
\begin{align}
	H_m(\psi) & = W^\ast \Lambda(\tau^{(m)}(\psi)) W,  \\
	\Lambda(\tau) & = \text{diag}(\gamma^0(\tau), \dots, \gamma^{N - 1}(\tau)), \\
\gamma^\ssi(\tau) & = \begin{cases}
	\exp(-2 \pi i \ssi \tau f_s / N) & \text{if }  \ssi <  \frac{N}{2}, \\
\cos(\tau \pi f_s )  & \text{if } \ssi = \frac{N}{2}, \\
	\exp(2 \pi i (N - \ssi) \tau f_s / N) & \text{if } \ssi >\frac{N}{2}. \\
\end{cases}
\end{align}
\end{subequations}
Here, $^\ast$ denotes the conjugate transpose operator. Further, $\tau^{(m)}(\psi)$ is the time shift of the signal at hydrophone $m$ caused by the \gls{DOA} $\psi$. Moreover, $W$ is the unitary discrete Fourier transform matrix, and $f_s$ is the sampling frequency of the hydrophone system.

The delay-and-sum beamformer reverses the delays in the signals, then
sums the signals, and finally calculates the energy of the summed signals.
That is, the beamformer output $B(\psi, \mathrm{y}_k)$ for \gls{DOA} angle $\psi$ and hydrophone sample batch $\mathrm{y}_k$ is given by
\begin{equation}
	B(\psi, \mathrm{y}_k) = \lVert \mathrm{H}^\top(\psi) \mathrm{y}_k \lVert_2^2,
\end{equation}

\subsection{Detection-Based Measurement Model}


Traditional target detection methods take target detections and associated bearings as inputs.
In passive sonar,
detections are obtained by applying a peak detector, typically a \gls{CFAR} detector, to
the beamformer output $B(\psi, \mathrm{y}_\ti)$. The resulting measurements are a set of bearings,
\begin{equation}
	z_\ti \equiv \{\psi_0, \dots, \psi_D\},
\end{equation}
where the number of detections $D$ will vary.
The cell-averaging \gls{CFAR} detector evaluates $B(\psi, \mathrm{y}_\ti)$ over a discrete set of
angles $\psi$. Each bearing-value pair $(\psi, B(\psi, \mathrm{y}_\ti))$,
known as the cell under test, is a detection candidate.
To account for the ambient noise variations, a Gaussian distribution is fitted to the energy in neighboring cells, including those from previous time steps. A significance threshold calculated from the fitted Gaussian distribution is then used to classify whether the cell under test deviates from the training cells. If so, the bearing of the test cell is added to the set of detections in $z_\ti$. 
More details on the \gls{CFAR} detector can be found in \cite{Richards2022}.

Among the true detections, there will be false detections. The false detections are
assumed to follow a Poisson point process, implying that the likelihood of observing $z_\ti$ when no target is present is given by
\begin{equation}
	\label{eq:background_clutter}
	\likelihood_0(z) = e^{-\lambda} \prod_{\psi \in z} \lambda \kappa(\psi).
\end{equation}
Here, $\lambda$ is the Poisson point process intensity, and $\kappa(\psi)$ is the \gls{PDF} of each false detection.
The false detections are assumed to be independent, identically, and uniformly distributed over the beamforming interval.

The likelihood of $z_\ti$ given that a target with state $x$ exists is
derived in \cite[Sec. V]{Ristic2013} and is given by
\newcommand{\pd}{p_\text{d}}
\begin{equation}
	\begin{split}
		\likelihood_1(z|x) =& \likelihood_0(z)(1 - \pd)  + \pd \sum_{\psi_d \in z} g(\psi_d|x)\likelihood_0(z \setminus {\psi_d}),
	\end{split}
\end{equation}
where $(\cdot \setminus \cdot)$ denote the set difference, $\pd$ is the probability that the target generates a detection, and $g(\psi_d|x)$ is
the likelihood function of detection $\psi_d$ due to the target. The measurement uncertainty is assumed
to follow a Gaussian distribution with variance $R$. That is,
\begin{equation}
	g(\psi_d|x) = \mathcal{N}(\psi_d; \psi, R),
\end{equation}
where $\psi$ is the bearing of the target.
In summary, the log-likelihood ratio of the detection measurements is given by
\begin{multline}
		\ln \likelihoodratio(z|x)  = \ln \likelihood_1(z|x) - \ln \likelihood_0(z) \\
	 	=  \ln \left( 1 - \pd + \pd \sum_{\psi \in z} \mathcal{N}(\psi_d; \psi, R)  \frac{\likelihood_0(z \setminus {\psi})}{\likelihood_0(z)} \right).
\end{multline}

A CFAR-based tracker is computationally efficient as it compresses samples into a set of detections. Since a significant amount of data is discarded in the detection process, a high SNR is required for the tracker to function well. Furthermore, the training cells may not accurately
represent the ambient noise, leading to a degraded performance. As will be shown next, this issue may be addressed by defining the
measurement model in terms of the raw hydrophone samples. The drawback is an increased computational complexity.

\subsection{Proposed \gls{TkBD} Measurement Model}

The proposed \gls{TkBD} measurement model assumes that each batch of hydrophone samples
follows a multivariate-t distribution
\begin{equation}
	\mathrm{y}_\ti \sim t_{N M}\left(\nu, 0, \frac{\nu - 2}{\nu}\Sigma(\psi_\ti)\right) \quad \nu > 2,
\end{equation}
where
\begin{equation}
	\Sigma(\psi) = \mathrm{H}(\psi) \Sigma_{ss} \mathrm{H}^\top(\psi) + \Sigma_{ee},
\end{equation}
$\Sigma_{ss} = \text{cov}(\mathrm{s}_\ti)$, and $\Sigma_{ee} = \text{cov}(\mathrm{e}_\ti)$.
Further, the \gls{PDF} of the multivariate-t distribution is given by
	\begin{multline}\label{eq:t_dist_pdf}
	  t_{NM}(\mathrm{y}; \nu, \mu, S)= \frac{\Gamma([\nu + NM]/2)}{\Gamma(\nu/2)(\nu \pi)^{NM / 2} |S|^{1/2} } \\
	  \cdot \left(
		1 + \frac{1}{\nu}((\mathrm{y}-\mu)^\top S^{-1} (\mathrm{y}-\mu))
	\right)^{-\frac{\nu + NM}{2}},
	\end{multline}
where $\Gamma$ is the gamma function, $\nu$ is the degrees of freedom, $\mu$ is the median, and $S$ is the scale matrix. For $\nu>2$ then $\text{cov}(\mathrm{y})=\frac{\nu}{\nu-2}S$. Noteworthy is that when $\nu\rightarrow\infty$, the frequently used Gaussian distribution is obtained.

To evaluate the \gls{PDF} in \eqref{eq:t_dist_pdf}, the determinant and the inverse
of $\Sigma(\psi_\ti)$ must be computed, which is computationally expensive due to its size $NM \times NM$. This computational complexity may be unmanageable for particle filter-based trackers that evaluate the measurement likelihood many times at each step of the filter recursions.  However, prior work \cite{Bosser2022} demonstrates that if the signal $\mathrm{s}_k$ and noise $\mathrm{e}_k$ are temporally and spatially white, i.e.,  $\Sigma_{ss} = \sigma_s^2 I_N$ and $\Sigma_{ee} = \sigma_e^2 I_{N M}$, so
that $\Sigma(\psi) = \sigma_s^2 \mathrm{H}(\psi) \mathrm{H}^\top(\psi) + \sigma_e^2 I_{NM}$, then
\begin{subequations}
	\begin{equation}
		\log |\Sigma(\psi)|  =  N\log(M \sigma_s^2 + \sigma_e^2) + N (M - 1)\log(\sigma_e^2),
	\end{equation}
	and
	\begin{equation}
		\mathrm{y}^\top \Sigma^{-1}(\psi) \mathrm{y} =  \frac{\lVert \mathrm{y} \rVert_2^2}{\sigma_e^2} - \frac{\sigma_s^2}{\sigma_e^2}\frac{B(\psi, \mathrm{y})}{\sigma_e^2 + M \sigma_s^2}.
	\end{equation}
\end{subequations}
Hence, under these assumptions, the log of the \gls{PDF} in \eqref{eq:t_dist_pdf} can be efficiently calculated from the beamformer $B(\psi, \mathrm{y})$ and the signal energy $\lVert \mathrm{y}\rVert^2$.

In reality, the ambient noise $\mathrm{e}_k$ is seldom white, i.e., $\Sigma_{ee}
\neq \sigma_e^2 I_{NM}$. However, if covariance of the ambient noise $\Sigma_{ee}$ is known the batch data $\mathrm{y}_k$ may be
whitened as
\begin{equation}
	\label{eq:naive_whitening}
		\tilde{\mathrm{y}}_\ti  = {\Sigma}_{ee}^{-\top/2} \mathrm{y}_\ti  \sim t_{NM}(\nu, 0, \bar{\Sigma}_{ss}(\psi) + I_{N M}),
\end{equation}
where
\begin{equation}
\bar{\Sigma}_{ss}(\psi) = {\Sigma}_{ee}^{-\top/2}\mathrm{H}(\psi) \Sigma_{ss} \mathrm{H}^\top(\psi) {\Sigma}_{ee}^{-1/2},
\end{equation}
Next, if one assumes that the source signal $\mathrm{s}_k$ is white so that $\Sigma_{ss}=\sigma^2_s I_N$ and neglecting the effect (except for scaling) of the whitening on the source signal, then
\begin{equation}
	\label{eq:big_approx}
\bar{\Sigma}_{ss}(\psi) \approx \eta \mathrm{H}(\psi) \mathrm{H}^\top(\psi).
\end{equation}
Here, $\eta$ is the power of the signal originating from the target relative to the ambient noise power after whitening, i.e., $\eta$ is the \gls{SNR} after whitening.
The motivation for making the ad-hoc approximation in \eqref{eq:big_approx} is that it enables the \gls{PDF} of $\tilde{\mathrm{y}}_k$ to be calculated efficiently via the beamformer. The argument in support of the approximation is that the distortion caused to the signal $\mathrm{s}_k$ by the whitening is likely negligible compared to other modeling errors, such that the assumption of $\mathrm{s}_k$ being a white signal; this assumption will be further discussed in Sec.~\ref{s:evaluation}.

In summary, the proposed measurement model uses the batch $\tilde{\mathrm{y}}_\ti$ of whitened hydrophone
samples as inputs, i.e.,
\begin{equation}
	z_\ti \equiv \tilde{\mathrm{y}}_\ti.
\end{equation}
The likelihood of $z$ given that target with state $x$ is modeled as
\begin{equation}
	\label{eq:meas_model1}
	\likelihood_1(z|x) = t_{NM}\left(z; \nu, 0, \frac{\nu - 2}{\nu}(\bar{\Sigma}_{ss}(\psi) + I_{NM})\right).
\end{equation}
When no target is present, i.e., $\eta=0$, the likelihood simplifies to
\begin{equation}
	\likelihood_0(z) = t_{NM}\left(z; \nu, 0, \frac{\nu - 2}{\nu} I_{NM}\right).
\end{equation}
Thus, the log-likelihood ratio can efficiently be calculated from the beamformer as
\begin{equation}
	\begin{split}
		\ln \likelihoodratio(z|x)  = & -\frac{N}{2}\ln\left( {M \eta + 1} \right) \\
			   & - \frac{\nu + NM}{2}\ln(1 - c B(\psi, z) ),
\end{split} \label{eq:likrat}
\end{equation}
where
\begin{equation}
	c = \frac{\eta}{(\nu  + \norm{z}^2) (1 + M\eta )}.
\end{equation}
For a derivation, see Appendix.

\subsection{Data-efficient Learning of $\Sigma_{ee}$ using \gls{VAR} Models}
Typically, the covariance $\Sigma_{ee}$ of the ambient noise is unknown and must be learned from historical data. Directly estimating $\Sigma_{ee}$ using the sample covariance matrix requires a substantial amount of data.
Instead, it is proposed that the correlation structure of the ambient noise is modeled using a \gls{VAR} model. This allows for more data-efficient learning of $\Sigma_{ee}$ and, as will be shown, the whitening to be done without factorizing and inverting $\Sigma_{ee}$. \gls{VAR} models have previously been successfully applied in other sonar applications to capture the spatial and temporal characteristics of underwater sounds
\cite{Barthelemy}, though they are more commonly used in economic modeling to capture relationships between variables over time \cite{Helmut2005}.

A $p$:th order \gls{VAR} model describes the ambient noise $\bm{e}_\ssi$ as
\begin{equation}\label{eq:ar_array}
	\bm{e}_\ssi  = A_1 \bm{e}_{\ssi - 1} + \dots + A_p \bm{e}_{\ssi - p} + \Sigma_{w}^{1/2} \bm{w}_\ssi,
\end{equation}
where $\bm{w}_\ssi$ is white noise with $\text{cov}(\bm{w}_\ssi)=I_M$. The matrices $A_1$, \dots, $A_p$, and $\Sigma_w$ are model parameters
that defines the structure of $\Sigma_{ee}$. However, instead of directly constructing $\Sigma_{ee}$  from the model parameters and then factorizing and
inverting the matrix to do the whitening in \eqref{eq:naive_whitening}, the fact that the \gls{VAR} model is invertible can be used. That is, the white noise $\bm{w}_\ssi$ can be retrieved from $\bm{e}_\ssi,\ldots,\bm{e}_{\ssi - p}$ as follows
\begin{equation}\label{eq:inverted VARA}
	w_\ssi = \Sigma_{w}^{-1/2}(\bm{e}_\ssi - A_1 \bm{e}_{\ssi - 1} - \dots - A_p \bm{e}_{\ssi - p}).
\end{equation}
Hence, the whitening of $\bm{y}_k$ in \eqref{eq:naive_whitening} can be efficiently implemented using the inverted \gls{VAR} model by substituting $\bm{e}_\ssi$ with $\bm{y}_k$ in \eqref{eq:inverted VARA}.


The model parameters $A_1, \dots, A_p, \Sigma_w$  can be learned using linear least squares \cite[Ch. 3]{Helmut2005}. That is, the parameters are calculated as
\begin{subequations}
\begin{equation}
  \{\hat{A}_1,\ldots,\hat{A}_p\}=\argmin_{A_1,\ldots,A_p} \sum_{\ssi=p+1}^{N_t} \epsilon^\top_\ssi \epsilon_\ssi
\end{equation}
where
\begin{equation}
	\epsilon_\ssi = \bm{y}_\ssi - A_1 \bm{y}_{\ssi - 1} - \dots - A_p \bm{y}_{\ssi - p},
\end{equation}
\end{subequations}
and $N_t$ denotes the number of samples in the dataset used to estimate the parameters. Further, the covariance $\Sigma_w$ is estimated as
\begin{equation}
  \hat{\Sigma}_w=\frac{1}{N_t-p-1}\sum_{\ssi=p+1}^{N_t} \epsilon_\ssi \epsilon^\top_\ssi.
\end{equation}
The model order $p$ can be selected using, e.g., the Akaike information criterion \cite[p.221]{LjungSysId}. An alternative method for learning the parameters that take into account the heavy-tailed distribution of the data can be found in \cite{Liu2019}.

\section{Target Dynamics Model}
\label{sec:target_dyn_mod}

Recall from Sec. \ref{sec:target_tracking} that the Bernoulli filter
recursions require a model of the target state dynamics $\pi(x|x')$ and a birth model $b_{\ti|\ti-1}(x)$.
These models are defined next.

\subsection{Motion Model}
A bearing-only target detection and tracking setup with only one hydrophone array is considered. To that end, let the target state at time step $\ti$ be
\begin{equation}
	x_\ti = \begin{bmatrix}\psi_\ti & \dot{\psi}_\ti & \eta_\ti^\dBB\end{bmatrix}^\top,
\end{equation}
where $\dot{\psi}_\ti$ is the bearing change rate and $\eta_\ti^\dBB$ is the SNR $\eta_\ti$ in dB, i.e., $\eta_\ti^\dBB = 10 \log_{10} \eta_\ti$. Changes in target bearing $\psi$ are modeled according to a constant velocity model, and the \gls{SNR} $\eta^\dBB$ of the target is modeled as a random walk. Details about these models and other motion models commonly used in target tracking can be found in \cite{RongLi2003}. Hence, conditioned on target state $x_\ti$, the \gls{PDF} of $x_{\ti + 1}$ is modeled as
\begin{equation}
	\pi(x_{\ti + 1}|x_{\ti}) = \mathcal{N}(x_{\ti + 1}; F x_{\ti}, G Q G^\top),
\end{equation}
where
\begin{gather}
	F = \begin{bmatrix}
		1 & T & 0 \\
		0 & 1 & 0 \\
		0 & 0 & 1
	\end{bmatrix}
	\
	G = \begin{bmatrix}
		T^2 / 2 & 0 \\
		T & 0 \\
		0 & T
	\end{bmatrix}\
	Q = \begin{bmatrix}
		q_\text{CV}^2 \!&\! 0 \\
		0\! &\!q_\text{dBSNR}^2
	\end{bmatrix}.\nonumber
\end{gather}
Here $T$ is the time between instant $\ti$ and $\ti+1$. Further, $q_\text{CV}^2$
and $q_\text{dBSNR}^2$ are the process noise variances for the constant velocity model
and the random walk model, respectively.

\subsection{Birth Model}
The birth model $b_{\ti|\ti - 1}(x)$ describes the probability distribution of the state $x$ of a new target given the latest measurement $z_{\ti - 1}$. Since $z_{\ti - 1}$ contains information about the bearing $\psi$ and the \gls{SNR} $\eta^\dBB$, they are assumed to be distributed proportionally to the likelihood ratio of the measurement. That is, their joint \gls{PDF} is modeled as
\begin{equation}
	\label{eq:birth_model_1}
	p(\psi, \eta^\dBB|z_{\ti-1}) \propto \likelihoodratio(z_{\ti-1}| x) p(\eta^\dBB),
\end{equation}
where $p(\eta^\dBB)$ denotes an uniform prior distribution assigned to the \gls{SNR} $\eta^\dBB$. 
The prior $p(\eta^\dBB)$ should reflect the expected \gls{SNR}
of yet to be detected targets.
This ensures the tracker is less prone to lock onto a noisy source that momentarily produces much sound energy.

The measurement $z_{\ti - 1}$ contains no information about the bearing change rate $\dot{\psi}$. Consequently, $\dot{\psi}$ is assumed to be normal distributed as $\mathcal{N}(\dot{\psi}; 0, P_{\dot{\psi}})$. Bringing it all together, the  probability distribution of the state $x$ of a new target is modeled as
\begin{equation}
	b_{\ti|\ti - 1}(x) = p(\psi, \eta^\dBB|z_{\ti-1}) \mathcal{N}(\dot{\psi}; 0, P_{\dot{\psi}}).
\end{equation}

\section{Evaluation}\label{s:evaluation}

\begin{figure*}[tb!]
	\centering
	\includegraphics[trim={0 0 3.5cm 0}, clip, width = 0.99\textwidth]{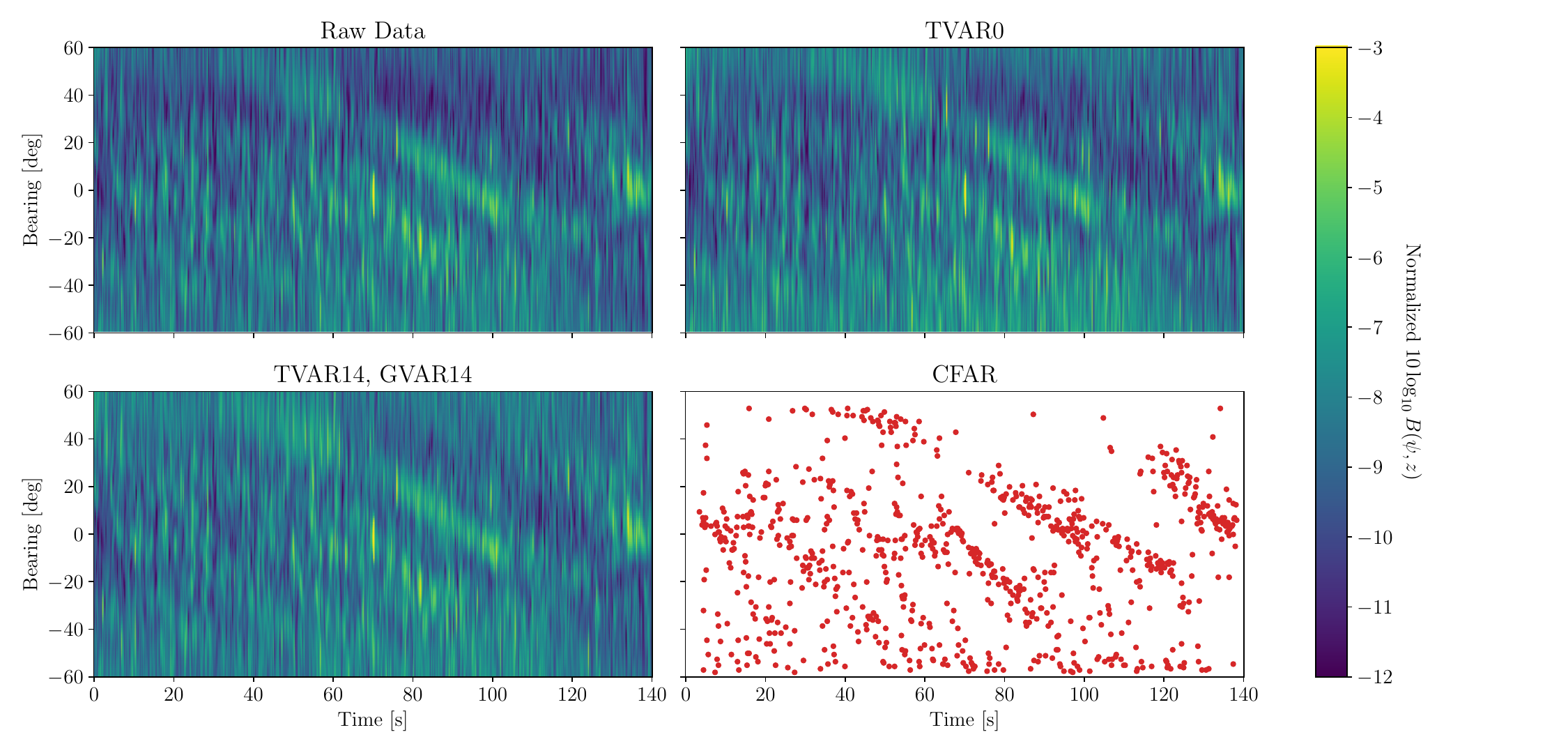}
	\caption{\label{fig:prepreproccesings}
		CFAR detections and BTRs after the data whitening step in the different trackers. All BTRs are normalized by a factor $b$ so that $\max_{\psi, \ti} bB(\psi, z_\ti) = 1$.
	The \gls{BTR} of the raw samples is shown in the top left. The CFAR detections are shown in the bottom right. The top right plot shows the \gls{BTR} of the whitened data in the \textit{TVAR0} tracker, which uses a \gls{VAR} model of order $p=0$. The bottom left plot shows the \gls{BTR} after the whitening in the \textit{GVAR14} and \textit{TVAR14} trackers, which uses a \gls{VAR} model of order $p=14$.}
\end{figure*}

The proposed tracking method is evaluated through experiments on both real and
simulated datasets. The objective of the evaluation is to examine the impact of
the ambient noise modeling and the associated data whitening, as well as the use of the t-distribution for modeling
the raw acoustic data. These factors are analyzed in isolation by comparing the proposed
tracker with variants that either do not include the ambient noise modeling and data whitening step or assume the data to be Gaussian
distributed. To that end, the following four trackers will be evaluated.
\begin{itemize}
	\item The \textit{TVAR14} tracker, which uses a \gls{VAR} model of order $p=14$ and a t-distributed signal model.
	\item The \textit{TVAR0} tracker, which uses a \gls{VAR} model of order $p = 0$  (equivalent to only modeling the spatial correlations in the noise) and a t-distributed signal model.
	\item The \textit{GVAR14} tracker,  which uses a \gls{VAR} model of order $p=14$ and a Gaussian distributed signal model.
	\item The \textit{CFAR} tracker, which uses detections from a CFAR detector as its input.
\end{itemize}
The CFAR detection based tracker is included as a reference. One may argue that the CFAR based tracker should include a whitening step similar to the other trackers. However, the CFAR tracker uses spatially and temporally distributed training cells, which should compensate for the spatiotemporal correlated ambient noise. Applying a whitening step would duplicate existing compensatory mechanisms.

\subsection{Performance Metrics}
The performance of each tracker is assessed by analyzing the estimated probability that a target exists and the \gls{OSPA} metric \cite{Schuhmacher2008}. In the case of single-target tracking, the \gls{OSPA} metric is given by
\begin{equation}
	\bar{d}_f^{(\rho)}(Y_{\ti|\ti}, Y_\ti) =
	\sum_{\hat{x} \in Y_{\ti|\ti}} d^{(\rho)}(\hat{x}, x_\ti) + \rho^f (1 - |Y_{\ti|\ti}|),
\end{equation}
where $x_\ti$ is the ground truth target state, $f$ is the norm order, and
$d^{(\rho)}$ is the distance measure with cut-off $\rho$. Here, targets are compared based
on their bearing differences, i.e.,
\begin{equation}
	d^{(\rho)}(\hat{x}, x_\ti) = \min (\lVert \hat{\psi} - \psi_\ti \rVert_f, \rho).
\end{equation}
Furthermore, $Y_{k|k}$ is the set of confirmed targets
\begin{equation}
	Y_{\ti|\ti} = \{x \, | \, x \in
X_{\ti|\ti} \text{ and } q_{\ti|\ti} > \gamma \},
\end{equation}
where $\gamma = \qty{90}{\percent}$.  The cut-off is set to be $\rho= \qty{30}{\degree}$ and the norm order $f = 1$ is used.

\subsection{Parameter Settings}

\begin{table}[tb!]
	\centering
	\caption{\label{tab:parameters}Parameter values used in the evaluation.
	Values in parentheses are values used in the simulation if they differ
from the parameter values used in the real-world data evaluation.}
\begin{tabular}{crl} \toprule
	Sym. & Value & Description \\ \midrule
	$p_\text{s}$ & $1 - 10^{-6}$ (0.99347) & Prob. of survival. \\
	$p_\text{b}$ & $2 \cdot 10^{-10}$ ($4.56 \cdot 10^{-8}$) & Prob. of target birth. \\
	$q_\text{CV}$ & $\qty{0.13}{\degree \per \second^2}$  & Motion process noise.  \\
	$q_\text{dBSNR}$ & $\qty{0.05}{dB \per \second}$  & SNR process noise.  \\
	$P_{\dot{\psi}}$ & \qty{0.001}{\degree^2 \per \second^2} & \makecell[l]{Initial uncertainty in target bearing \\ change rate} \\
	$\nu$ & 3 (12) & Deg. of freedom in mv-t dist. \\
	$N$ & 64 & Number of samples per batch. \\
	$T$ & \qty{0.17}{\second}  & Time between batches. \\
	$p$ & 14 & Model order of the \gls{VAR} model.  \\
	\bottomrule
\end{tabular}
\end{table}

The uniform prior $p(\eta^\dBB)$ on the SNR for newborn states $x$ in \eqref{eq:birth_model_1} is calibrated individually for each tracker. The lower and upper bounds of the prior are increased simultaneously until false tracks start to appear in a target-free dataset. This ensures that each tracker is as sensitive as possible without generating false detections when applied to the target-free dataset. For the CFAR-based tracker, the false detection intensity $\lambda$ in \eqref{eq:background_clutter} is adjusted similarly. All other parameters are kept the same to the greatest extent possible to ensure a fair comparison of the trackers. The parameter values used are listed in Tab.~\ref{tab:parameters}.

\subsection{Simulated Scenario}
\begin{figure}[tb!]
	\centering
	\subfloat[Simulated scenario (not to scale). The target begins its journey in the upper part of the figure, traveling in the negative $y$ direction. The eight dots indicate the array. ]{\includegraphics[width = 0.45\linewidth,valign=c]{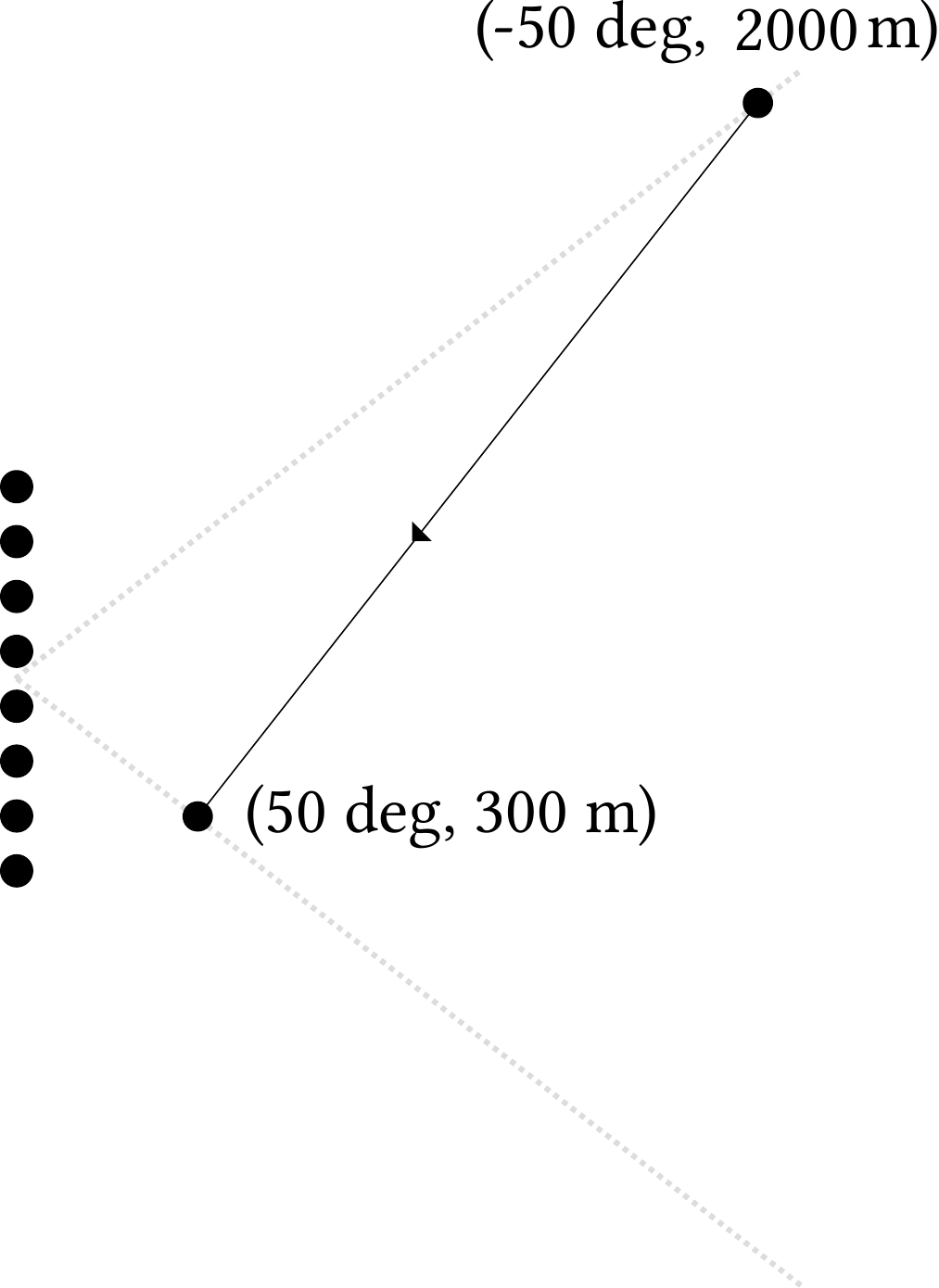}
     \vphantom{\includegraphics[width= 0.45\linewidth,valign=c]{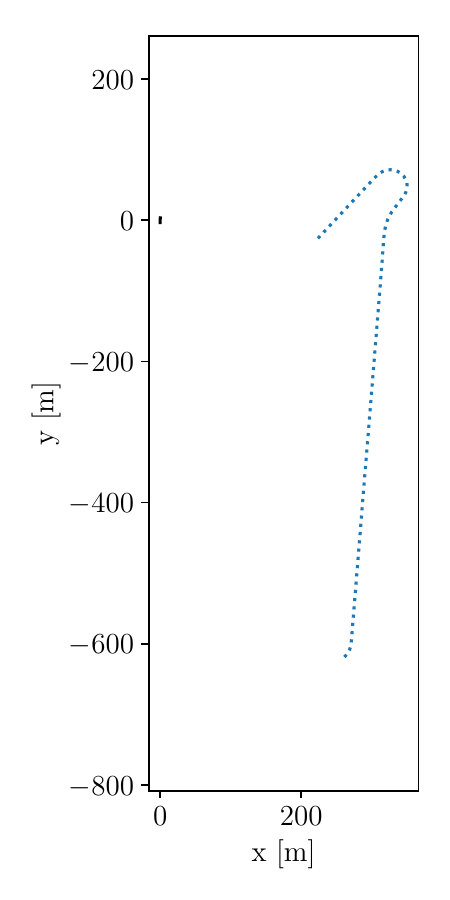}}}
	\hfill
	\subfloat[Real-world scenario. The array is located at $(0,0)$, and the target begins its journey in the lower part of the figure, traveling in the positive $y$ direction.]{\includegraphics[width = 0.45\linewidth,valign=c]{figures/cartesian_real.pdf}}
	\caption{\label{fig:overview} Illustration of the target trajectories and array locations in the simulation and real-world scenarios.}
\end{figure}

The simulated scenario consists of a target that moves towards an 8-element array tuned for $\qty{800}{\hertz}$ signals. 
The frequency content of the considered signals is between \qty{750}{\hertz} and \qty{937.5}{\hertz}, demodulated using a \qty{750}{\hertz} cosine signal and sampled at a sampling frequency of \qty{375}{\hertz}. 
The target moves at a constant velocity of \qty{2.5}{m/s}, starting at a bearing of $\ang{-50}$ at a distance of \qty{2000}{m} away from the array and ending at bearing \ang{50} at a distance of \qty{300}{m} away from the array.
The trajectory is shown in Fig. \ref{fig:overview}a. The range to the target is mapped to an \gls{SNR} according to
\begin{equation}
	\eta^\dBB = 10 \log_{10} \left( \frac{r}{200} \right)^{1.8},
\end{equation}
where $r$ is measured in meters. This mimics a propagation loss between
cylindrical and spherical spreading~\cite[p. 39]{fmv2014}. Generating the
simulated dataset is a multistep process, as detailed in Alg.~\ref{alg:two}.
Notably, the target signal is
added to the correlated noise, which means that the inaccuracies of the approximation in \eqref{eq:big_approx} should affect
the trackers similarly as it will in the later analyzed real data scenario. The same \gls{VAR} model parameters are used both in the data generation and the data whitening in the tracker.

\SetKwComment{Comment}{/* }{ */}

\begin{algorithm}[tb!]
    \SetKwInOut{Input}{input}
    \SetKwInOut{Output}{output}
\caption{Generating a simulated dataset.}\label{alg:two}
\Input{Target bearing $\psi_\ti$ and \gls{SNR} $\eta_\ti$, \gls{VAR} parameters $A_1, \dots, A_p$, $\Sigma_w$, and degrees of freedom $\nu$ of the t-distribution.}
\Output{Sample $\mathrm{y}_\ti$}

\Comment{Sample ambient signal}
Sample $\bm{w}_n \sim \mathcal{N}(0, I_M)$

Compute $\bm{e}_n$ using \eqref{eq:ar_array} and $\bm{w}_n$

Estimate $\sigma_e^2 = |\expect{[\bm{e}_n \bm{e}_n^\top ]}|^{1/M}$

Batch $\bm{e}_n$ into $\mathrm{e}_\ti$ as in \eqref{eq:e_batch}

\Comment{Sample target signal}
Calculate $\sigma_s^2 = \eta_\ti \sigma_e^2$

Sample $\mathbf{s}_\ti \sim \mathcal{N}(0, \sigma_s^2 \mathrm{H}(\psi_\ti) \mathrm{H}^\top(\psi_\ti))$

\Comment{Construct the full multivariate-t distributed signal}

Sample $c_\ti \sim \chi^2(\nu)$

Compute $\mathrm{y}_\ti = \sqrt{\nu / c_\ti} \cdot (\mathrm{s}_\ti + \mathrm{e}_\ti )$
\end{algorithm}

\begin{figure}[t!]
	\centering
	\includegraphics[trim={0.3cm 0.5cm 0.3cm  0.3cm},clip,width =\linewidth]{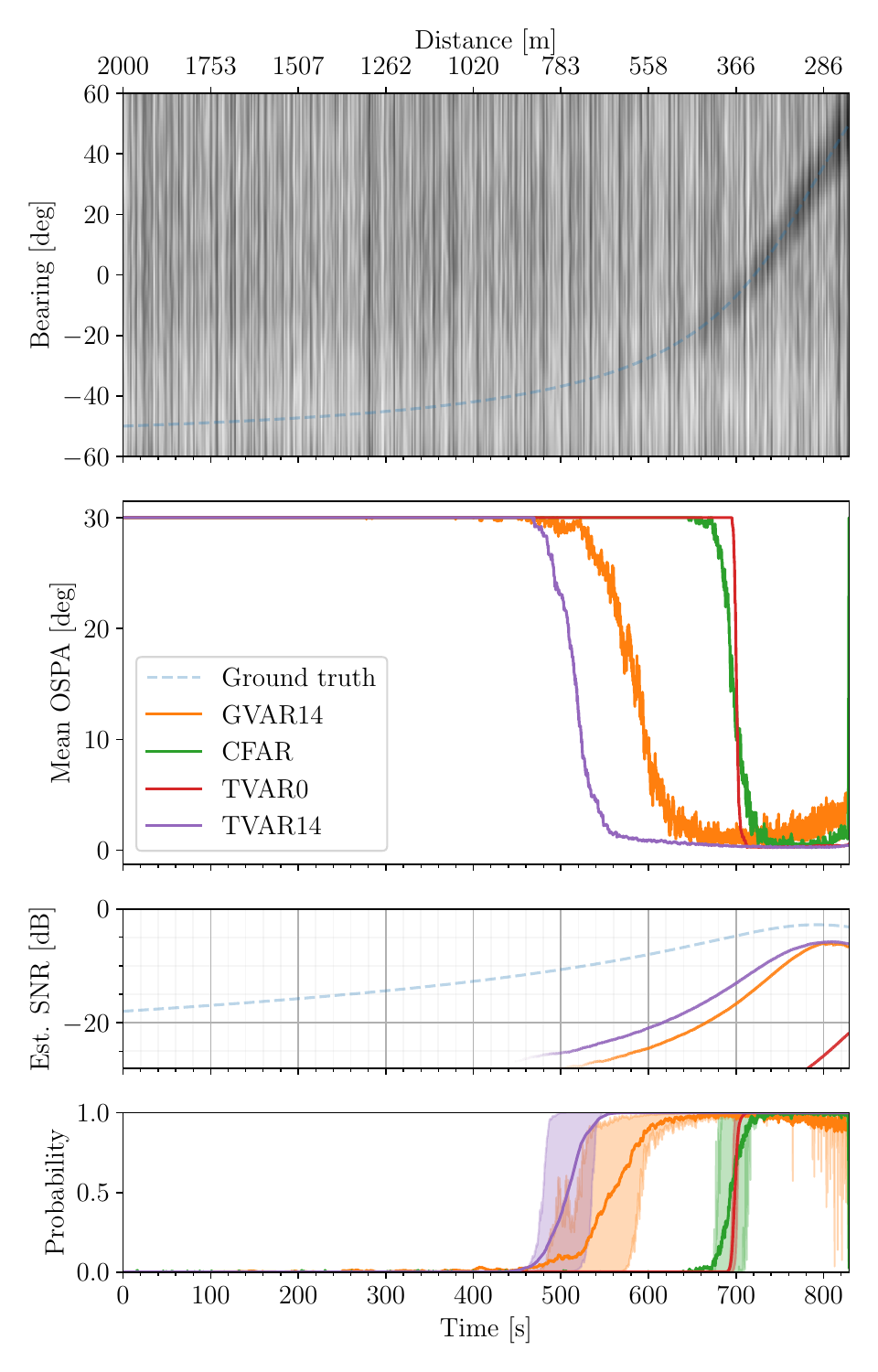}
	\caption{\label{fig:sim-results}
Results of 100 Monte Carlo runs of the simulated scenario. The top plot shows the \gls{BTR} of the generated signal for a single run, together with the ground truth bearing in dashed blue. Below the \gls{BTR}, the \gls{OSPA}, estimated \gls{SNR}, and estimated target existence probability are shown. The transparent regions in the estimated target existence probability correspond to the top 90 and bottom 10 quantiles.}
\end{figure}
\begin{figure}[t!]
	\centering
	\includegraphics[width = \linewidth]{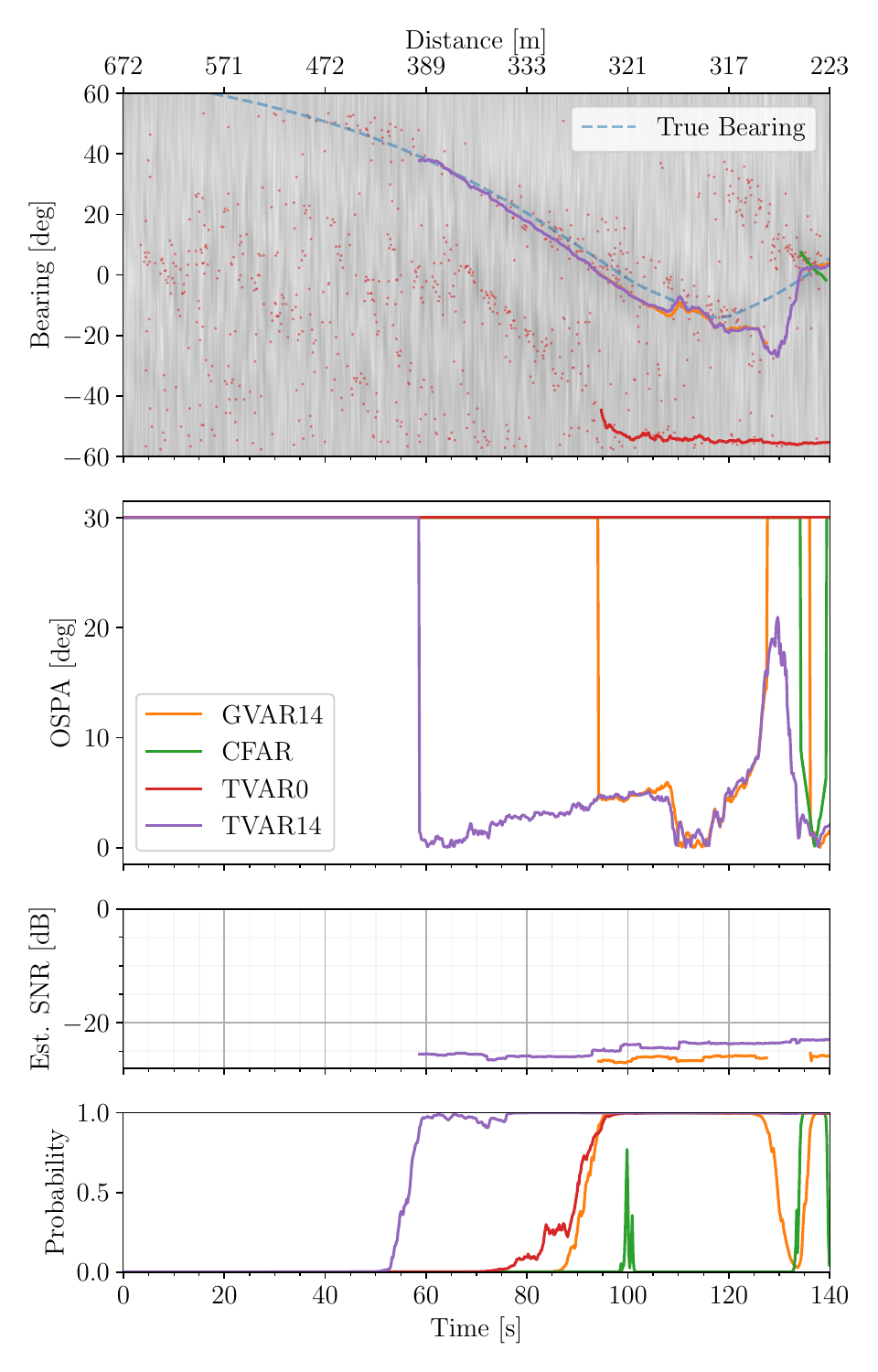}
	\caption{\label{fig:real-results}
Results from the real-world scenario. The top plot shows the \gls{BTR}, along with the truth and estimated bearings. Below the \gls{BTR}, the \gls{OSPA}, estimated \gls{SNR}, and estimated target existence probability are shown.}
\end{figure}

The true bearing, true SNR, $B(\psi, \mathrm{y})$, and the estimated tracks can be seen in Fig. \ref{fig:sim-results}. The results show that the CFAR tracker detects the target at approximately $\qty{700}{\second}$ when the target is around $\qty{300}{\meter}$ from the array. This corresponds to an \gls{SNR} of \qty{-5}{dB}. The \textit{TVAR0} tracker shows similar performance. Modeling the ambient noise using \gls{VAR} model of order 14 improves the detection performance, as demonstrated by the \textit{GVAR14} tracker. The \textit{GVAR14} tracker detects the target at a distance of \qty{550}{\meter}, which corresponds to an \gls{SNR}
of \qty{-8}{dB}. However, the estimated target existence probability fluctuates rapidly above and below the detection threshold, causing the \gls{OSPA} metric to vary correspondingly. Further improvement is observed when the Gaussian distribution is replaced by the proposed multivariate-t distribution, as seen from the performance of the \textit{TVAR14} tracker. The multivariate-t distribution stabilizes the estimates of target existence probability and leads to a smoother \gls{OSPA} measure. It also enhances the tracker's detection capability, enabling even earlier target detection. The target is detected at a distance of \qty{560}{\meter}, which corresponds to an SNR of \qty{-9}{dB}. It is noted that the \gls{SNR} is systematically underestimated by the trackers, which may result from the approximations made in the derivation of the measurement model.

\subsection{Real-world Scenario}
The real-world dataset was collected during a sea trial in the Stockholm archipelago using an 8-element horizontal hydrophone array and a SAAB AUV62 autonomous underwater vehicle acting as the target. The array's shape, orientation, and location were calibrated using the method described in \cite{Skog2024}.  The distance between the hydrophone elements was approximately \qty{0.93}{\meter}. Given that the speed of sound in the baltic sea is approximately \qty{1500}{\meter \per \second}, this corresponds to a design frequency of $\qty{800}{\hertz}$. The SAAB AUV62 followed the trajectory shown in Fig. \ref{fig:overview} (b), starting at a distance of \qty{675}{\meter} and ending at \qty{220}{\meter} from the array. Throughout the trajectory, the AUV maintained a speed of \qty{6.2}{\meter \per \second} (approximately \qty{12}{knots}) and a constant depth of \qty{25}{\meter}. The depth ranges from \qtyrange{30}{45}{\meter} at the test site, with a few islands located to the left of the array in the plot. Due to the shallow water at the site, the hydroacoustic environment is expected to be complex, with many reflections and other unmodeled properties. As a result, the t-distribution model in the \textit{TVAR14} and \textit{TVAR0} trackers were set to have a degree-of-freedom $\nu = 3$, the lowest value for which the covariance matrix of the t-distribution remains defined.

Before the trackers processed the recorded hydrophone, the data was preprocessed as follows. Similarly to the simulated scenario, the data were bandpass-filtered with cut-off frequencies \qty{750}{\hertz} and \qty{937.5}{\hertz}, then demodulated using a \qty{750}{\hertz} cosine signal and finally downsampled to a sampling frequency of \qty{375}{\hertz}.  In the frequency band \qty{750}{\hertz} and \qty{937.5}{\hertz}, the received signal can be considered approximately white. This is because the spectrum of the sound generated by the AUV62 is approximately flat, and the differences in propagation loss across the frequency band are negligible.

\subsubsection{Effects of Ambient Noise Modeling}
In Fig.~\ref{fig:prepreproccesings}, the effect of whitening the hydrophone data in Fig.~\ref{fig:btr-early}b with a \gls{VAR} model learned on the data in Fig.~\ref{fig:btr-early}a, is shown in terms of \gls{BTR} after the whitening process. Also shown are the detections found by applying the \gls{CFAR} detector to the same dataset. Note that the \textit{TVAR14} and \textit{GVAR14} trackers use the same \gls{VAR} model of order $p=14$ for the ambient noise modeling and data whitening, whereas the \textit{TVAR0} tracker uses a $p=0$ order model. The effects of the whitening processes are apparent by comparing the BTR after the data whitening in the \textit{TVAR14} and \textit{TVAR0} trackers to the BTR calculated on the raw data. The energy distribution across the bearings is more uniform after the whitening. Specifically, bearings within \qtyrange{20}{60}{\degree} exhibit lower signal energy in the \gls{BTR} of the raw data and in \textit{TVAR0}, compared to \textit{TVAR14} and \textit{GVAR14}. Conversely, the energy in bearings \qtyrange{-60}{-20}{\degree} during the timespan \qtyrange{90}{110}{\second} is higher in \textit{TVAR0}

\subsubsection{Detection and Tracking Performance}
In Fig.~\ref{fig:real-results}, the estimated target track, \gls{SNR}, target existence probability, and the calculated \gls{OSPA} for the four trackers are shown. The \gls{CFAR} tracker detects and starts to track the target at a distance of \qty{250}{\meter}. The \textit{TVAR0} tracker detects the target at a distance of \qty{325}{\meter}, but the estimated bearing is too far from the ground truth to be considered a valid track. This is in line with the observations in Fig.~\ref{fig:prepreproccesings}, where the negative bearings exhibited higher beamforming energy due to the absence of temporal whitening. The \textit{GVAR14} tracker detects the target at approximately \qty{325}{\meter} and maintains the track for approximately \qty{125}{\second} until the SNR temporarily decreases. However, the method quickly recovers the track.

A common issue across the CFAR, \textit{GVAR14}, and \textit{TVAR0} trackers is that they either lose track or initiate false tracks. In contrast, the \textit{TVAR14} tracker detects the target the earliest, when the target is \qty{390}{\meter} from the array, and maintains a stable track thereafter.

\section{Discussion and Conclusions}

The challenge of reliable broadband passive sonar target detection and tracking in complex acoustic environments has been addressed. A solution has been proposed based on a vector-autoregressive model for the ambient noise and a heavy-tailed statistical model for the distribution of the raw hydrophone data. These models have been integrated into a Bernoulli track-before-detect (TkBD) filter to realize a bearing-only tracker. To facilitate a computationally less expensive evaluation of the proposed statistical model, approximations have been introduced to compensate for the effects of the ambient noise via a recursive preprocessing step where the data is whitened. The whitening of the data facilitates the statistical model to be expressed as a function of the conventional beamformer, significanly reducing the computational complexity of the statistical model.  

The proposed solution has been evaluated on both simulated and real-world data. Results showed that the proposed vector-autoregressive model can learn and compensate for a lot of the spatiotemporal correlations in ambient noise. Further, results showed that the proposed heavy-tailed multivariate-t distribution model made the trackers more robust than the case when the data was modeled as Gaussian distributed. The simulations show that the SNR at which the target can be detected is reduced by \qty{4}{\dB} compared to when using the standard constant false alarm rate detector based tracker. Further, the test with real-world data shows that the proposed solution increases the target detection distance from \qty{250}{\meter} to  \qty{390}{\meter}.

The presented results illustrate that the TkBD technology, in combination with data-driven ambient noise modeling and heavy-tailed statistical signal models, can enable reliable broadband passive sonar target detection and tracking in complex acoustic environments and lower the SNR required to detect and track targets.	Hence, the technology can contribute to more effective monitoring of critical underwater infrastructure and potentially increase the safety of maritime activities.

Future research should explore the possibility to integrate the proposed ambient noise model and heavy-tailed signal model in multi-target tracking. While multiple target \gls{TkBD} is still in development, methods such as the information exchange filter \cite{Davies2024} and belief propagation for multi-Bernoulli filters \cite{Liang2023} have shown promising results. The reason behind the observed bias in the SNR estimates should also be investigated.

\appendix

Given the \gls{PDF} in \eqref{eq:t_dist_pdf}, it holds that
\begin{align*}
	\begin{split}
		\likelihood_0(z)  & = t_{NM}\left(z; \nu, 0, \frac{\nu - 2}{\nu}I_{NM}\right) \\
			   & =  \frac{C}{|I_{NM}|^{1/2}} \left( 1 + \frac{\lVert z \rVert^2}{\nu - 2}  \right)^{-(\nu + NM) / 2},
	\end{split}
\end{align*}
for some constant $C$ and
\begin{align*}
	\begin{split} \likelihood_1(z|x)  & = t_{NM}\left(z; \nu, 0, \frac{\nu - 2}{\nu}\Sigma\right) \\
	& = \frac{C}{|\Sigma|^{1/2}} \left[ 1 + \frac{z^\top \Sigma^{-1} z}{\nu - 2}  \right]^{-(\nu + NM) / 2},
	\end{split}
\end{align*}
where $\Sigma = \eta \mathrm{H}(\psi) \mathrm{H}^\top(\psi) + I_{NM}$.
According to \cite{Bosser2022, Bosser2023}, it holds that
\begin{equation*}
	z^\top \Sigma^{-1} z  \approx  \norm{z}^2 - \frac{\eta B(\psi, z)}{(1 + M\eta)} ,
\end{equation*}
and
\begin{equation*}
	\ln |\Sigma| \approx   N \ln(M \eta + 1).
\end{equation*}
This gives the likelihood ratio
\begin{align*}
	\ln \likelihoodratio(z|x)  = & \ln \likelihood_1(z|x) - \ln \likelihood_0 (z) \\
	   & \approx - \frac{N}{2}\ln\left( {M \eta + 1} \right)  \\
	  & + \frac{\nu + NM}{2}  \left( \ln\left( 1+ \frac{\norm{z}^2}{\nu } \right)\right.  \\
	  & \left. - \ln \left( 1 + \frac{\norm{z}^2}{\nu } - \frac{\eta B(\psi, z)}{\nu (1 + M\eta )} \right) \right),
\end{align*}
By using that $\ln (a) - \ln(a -b) = - \ln(1 - b / a)$, the likelihood may be rewritten as
\begin{align*}
	\begin{split}
		\ln \likelihoodratio(z|x) \approx & - \frac{N}{2}\ln\left( {M \eta + 1} \right) \\
			   & - \frac{\nu + NM}{2}\ln\left(1 - c B(\psi, z) \right),
	\end{split}
\end{align*}
where
\begin{equation*}
	c = \frac{\eta}{(\nu  + \norm{z}^2) (1 + M\eta )}.
\end{equation*}

%
%
%
%
%
%
%
%
%
%

\bibliographystyle{IEEEtran}
\bibliography{IEEEabrv,library.bib}
\begin{IEEEbiography}[{\includegraphics[width=1in,height=1.25in,clip,keepaspectratio]{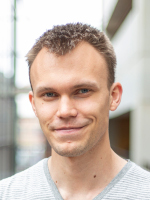}}]{Daniel Bossér} (Student Member IEEE) received his BSc and MSc degrees in Engineering Physics and in Engineering Mathematics and Computational Science from Chalmers University of Technology, Gothenburg, Sweden, in 2018 and 2020, respectively. He is currently a Ph.D. student Linköping University. His main research interests are 
	stochastic signal processing and sensor fusion with applications to underwater target tracking. 
\end{IEEEbiography}

\begin{IEEEbiography}[{\includegraphics[width=1in,height=1.25in,clip,keepaspectratio]{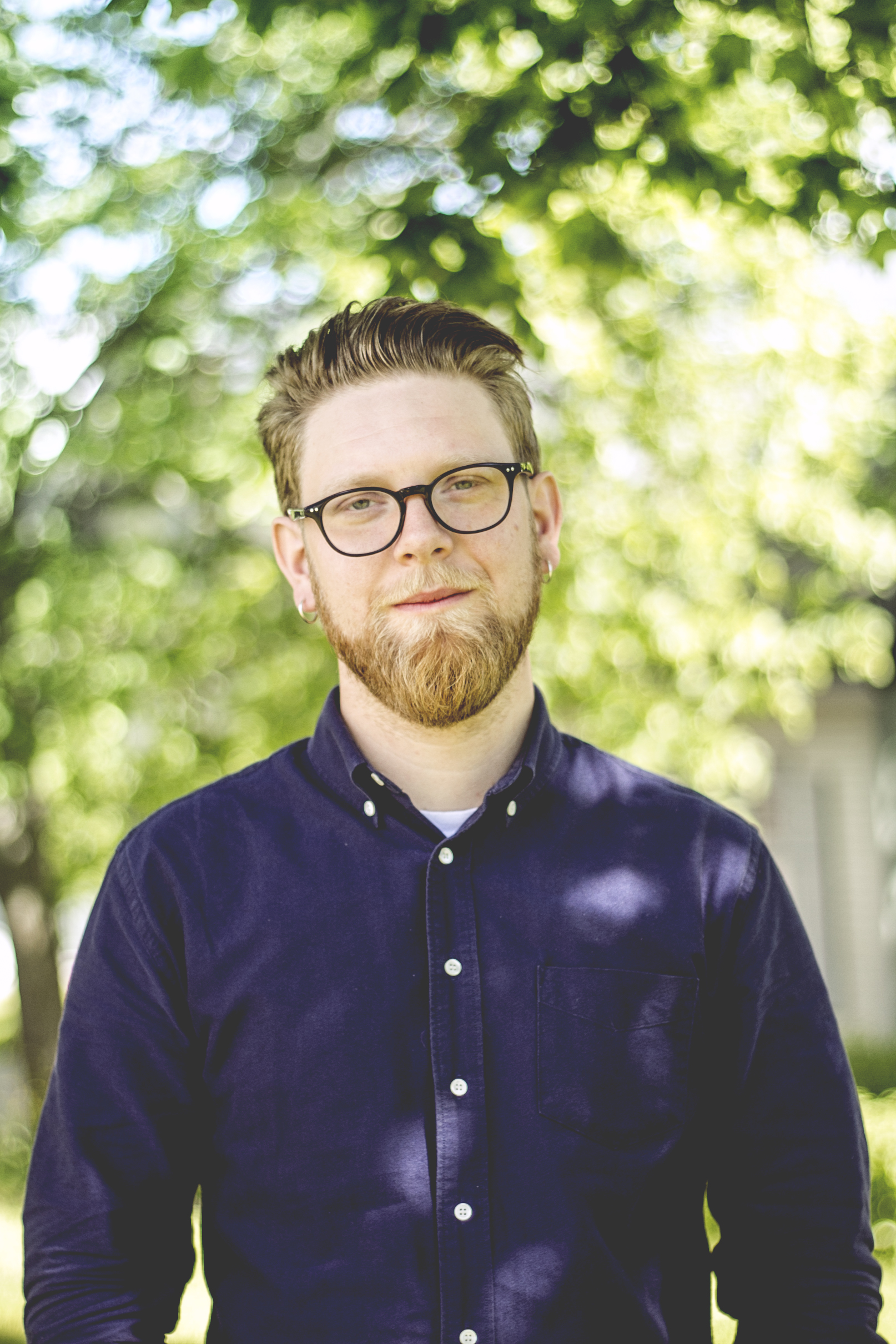}}]{Isaac Skog}(S'09-M'10) received his BSc and MSc degrees in Electrical Engineering from KTH Royal Institute of Technology, Stockholm, Sweden in 2003 and 2005, respectively.
In 2010, he received a PhD degree in Signal Processing with a thesis on low-cost navigation systems. In 2009, he spent 5 months with the Mobile Multi-Sensor System research team, University of Calgary,
Canada, as a visiting scholar and in 2011 he spent 4 months at the Indian Institute of Science (IISc), Bangalore, India, as a visiting scholar. Between 2010 and 2017 he was a
researcher at KTH Royal Institute of Technology, Stockholm, Sweden. Between 2017 and 2023 he was a
assistent professor in automatic control at Link\"{o}ping University, Sweden. Currently he is an associate professor in communication systems at KTH Royal Institute of Technology, Stockholm, Sweden,
and a senior researcher at the Swedish Defence Research Agency (FOI), Stockholm, Sweden. 
\end{IEEEbiography}

\begin{IEEEbiography}[{\includegraphics[width=1in,height=1.25in,clip,keepaspectratio]{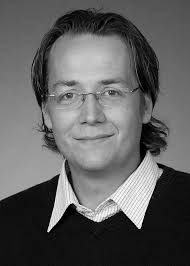}}]
{Magnus Lundberg Nordenvaad} was born in Lule\aa, Sweden, in 1973. He received the MSc degree in Computer Engineering from the Lule\aa~ University of Technology, Lule\aa, Sweden, in 1998, and the PhD degree in Signal Processing from the Chalmers University of Technology, Gothenburg, Sweden, in 2003. He has authored around 60 papers in international journals and conferences. His research interests lie in statistical signal processing and how it applies to sonar, digital communication, radar, and process diagnostics.

He has held faculty positions with the Department of Information Technology, Uppsala University, Uppsala, Sweden, and the Department of Computer Science and Electrical Engineering, Lule\aa~University of Technology. He has also held visiting researcher/professor positions at Purdue University, Colorado State University, and the University of Florida. He is currently a Deputy Research Director with the Swedish Defense Research Agency.
\end{IEEEbiography}

\begin{IEEEbiography}[{\includegraphics[width=1in,height=1.25in,clip,keepaspectratio]{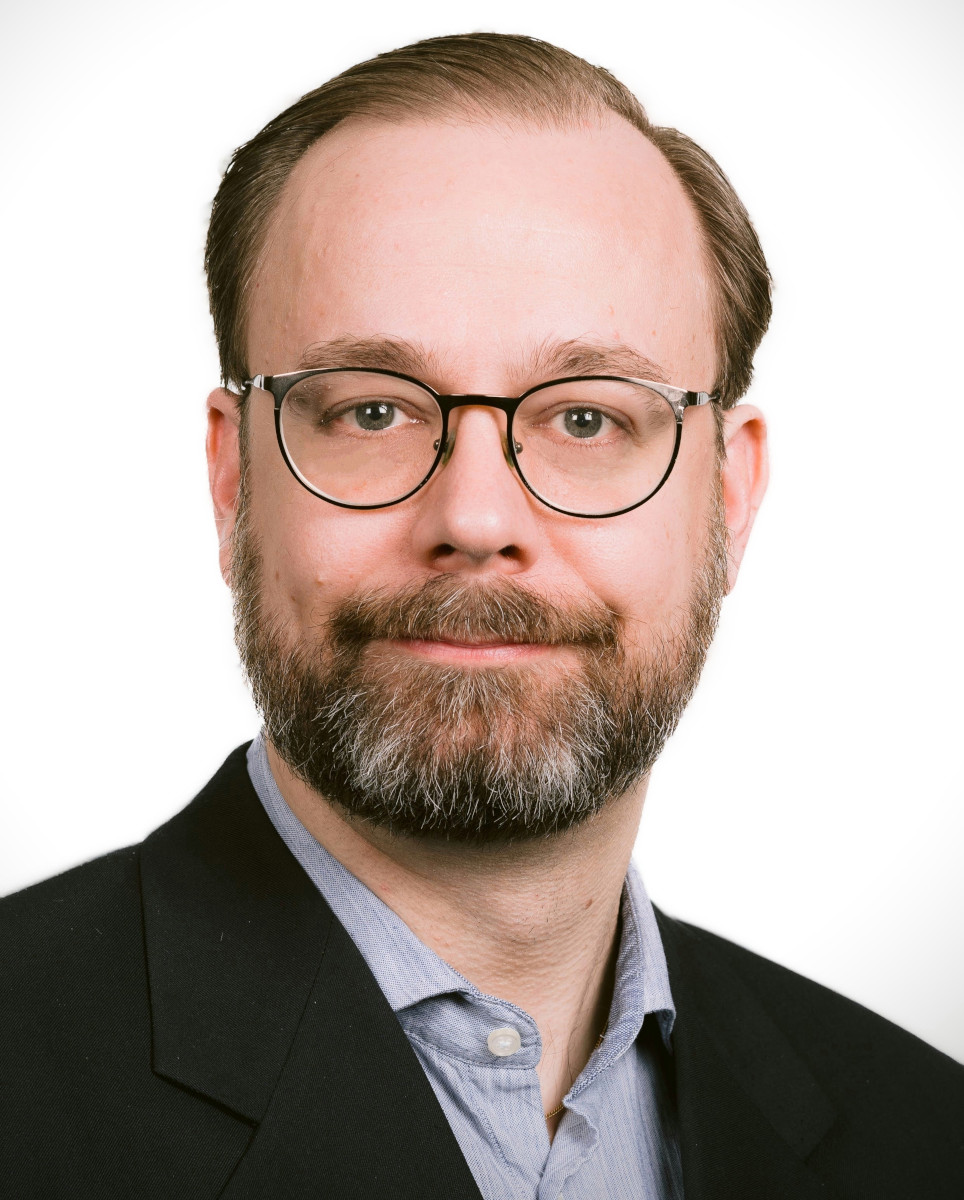}}]
{Gustaf Hendeby} (S'04-M'09-SM'17) received the  M.Sc.\ degree in applied physics and electrical engineering in 2002 and the Ph.D.\ degree in automatic control in 2008, both from Linköping University, Linköping, Sweden.

	He is Associate Professor and Docent at the division of Automatic Control, Department of Electrical Engineering, Linköping University.
  He worked as Senior Researcher at the German Research Center for  Artificial Intelligence (DFKI) 2009--2011, as Senior Scientist at Swedish Defense Research Agency (FOI) and held an adjunct Associate Professor position at Linköping University 2011--2015.
  His main research interests are sensor fusion and stochastic signal processing with applications to nonlinear problems, target tracking, and simultaneous localization and mapping (SLAM), and he is the author of several published articles and conference papers in the area.
  He has experience of both theoretical analysis as well as implementation aspects.

	Dr.\ Hendeby is since 2018 an Associate Editor for IEEE Transactions on Aerospace and Electronic Systems in the area of target tracking
  and multisensor systems, since 2024 Associate Editor in Chief for Journal of Advances in Information Fusion (JAIF), and since 2021 leader of the WASP Localization and Navigation Area Cluster.
  In 2022 he served as general chair for the 25th IEEE International Conference on Information Fusion (FUSION) in Linköping, Sweden and Technical Chair for the 26th and 27th IEEE International Conference on Information Fusion in Charleston, SC and Venice, Italy, respectively.
  He was elected into the International Society of Information Fusion (ISIF) board of directors for the term 2023--2025.
\end{IEEEbiography}
\end{document}